# Wavelength-tunable doughnut beam generation using few-mode long-period fiber grating

Kuei-Huei Lin and Wen-Hsuan Kuan*

*Department of Applied Physics and Chemistry, University of Taipei, Taipei, Taiwan*

E-mail: khlin@utaipei.edu.tw, wenhsuan.kuan@gmail.com

Ultra-broadband long-period fiber grating with 10-dB bandwidth of 155 nm and 3-dB bandwidth of 415 nm is demonstrated around 1060 nm, which has been combined with a tunable ytterbium-doped fiber laser to generate optical doughnut beams in the spectral range from 1030 nm to 1099 nm, and the vector nature of generated optical fields are observed. The long-period fiber grating has also been used to transform the amplified spontaneous emission of an ytterbium-doped fiber amplifier into a broadband optical doughnut beam.

Light-matter interaction can be controlled by manipulating the beam polarization, spatial intensity, and phase profile. In general, the transverse polarization distribution of a conventional laser beam is uniform with planar or spherical wavefronts, which we now call it as the scalar beam. However, recent studies confirm the existence of vector optical beams. The vector beam refers to the light with non-uniform transverse polarization distributions. One kind of the vector beams is a light beam whose polarization state is axially symmetric in the cross section, i.e., a cylindrical vector beam (CVB). The cylindrical vector beam is the characteristic solution of the paraxial Helmholtz equation in the cylindrical coordinate system. The polarization state of CVB is indeterminate at the propagation axis, where a polarization singularity exists. Among the various unconventional optical beams, the optical vortices are beams of high-order modes that carry non-zero orbital angular momentum (OAM) and are characterized by phase singularities and intensity singularities on the density profiles. When a spiral phase is added to the vector beam, the beam will acquire optical angular momentum and is termed as a vortex vector beam (VVB). In their lowest orders, these beams (CVB, OAM, and VVB) are usually characterized by doughnut-shaped beam profiles and have found various applications.[1]

Typical methods for doughnut beam generation use bulk optical devices, but they will usually encounter stability issues. For example, the generation of cylindrical vector beam can be carried out by interference, spatial light modulator (SLM) wavefront reconstruction, and crystal birefringence. However, these methods require complex optical systems, which are costly and depend on precise optical alignments. The entire system is bulky and its stability is easily influenced by the environment. It is also demonstrated that an offset coupling between SMF and MMF allows the generation of radially, azimuthally, or hybridly polarized light beams, but the coupling loss is usually very large.[2] An all-fiber method with reduced optical loss is thus appealing in the generation of vectorial light fields. Zhou et al. use a few-mode fiber Bragg grating (FM-FBG) as the mode converter in a passively mode-locked erbium-doped fiber laser.[3] Compared with bulk optics technology, the method of producing cylindrical vector beams with all-fiber components has higher stability and the system architecture is simpler. However, in order to suppress the $LP_{01}$ fundamental mode in the output beam to obtain a higher $LP_{11}$ mode purity, a narrow bandwidth FM-FBG is used in this erbium-doped fiber laser, such that the output spectral width of the mode-locked pulses is only 0.3 nm.

A fiber grating is a periodic perturbation in the refractive index along the fiber by using the photosensitivity of certain types of optical fibers. The formation of permanent gratings in an optical fiber was first demonstrated by Hill et. al, in 1978.[4] Fiber gratings can be classified into two major types: fiber Bragg gratings (FBGs) and long-period gratings (LPGs). Selected by Bragg reflection, the FBG will couple some specific core modes to the counter propagating modes. The devices of fiber Bragg grating have a period of about 1 μm and play a very important role in fiber optic communication systems and sensor technologies. The major applications are such as filtering, add/drop multiplexing, gain equalization, laser wavelength stabilization, and dispersion compensation for accumulated dispersion in the system. On the other hand, a LPG is a fiber device with a typical period of several hundred micrometers. It can be regarded as a transmission grating that couples the fundamental core mode to a series of co-propagation higher-order modes or cladding modes, and thus generates rejection bands in the transmission spectrum of the fundamental mode. The applications of LPGs include band rejection filters, gain flattening of erbium-doped fiber amplifiers (EDFAs), sensor technologies, and dispersion compensation, etc.[5–7] Conventionally, the fabrication of LPGs in usual fibers are proceeded via electric arc discharging, ultraviolet exposure, $CO_2$ laser inscription, or mechanical stress.[8–11] We have reported a method for generating both broadband chirped LPGs and constant-period LPGs in a bent PCF by using a constant-period V-grooved plate.[12] The central wavelengths and rejection bandwidths of LPGs could be tuned over a broad spectral range with adjustable transmission loss. Mechanically induced LPGs of 12 ~ 33 mm grating length showed resonant wavelength tuning range over 800 nm, and the 3-dB bandwidth can be controlled from 10 nm to 250 nm. However, the curvature-tuning of the bended fiber is difficult and the repeatability is not good.

Based on the coupling of fundamental core mode to a series of co-propagating higher-order modes, long-period fiber gratings (LPGs) are suitable as an all-fiber device for the generation of vector beams or OAMs.[13,14] The resonance wavelength depends on both the grating period and the effective indices of the coupling modes. However, the typical bandwidth of LPG is about 10 nm, which is insufficient for some applications.

In this work, we will make ultra-broadband LPGs in a few-mode fiber (FMF) with periodic mechanical stress on the fiber. By the adjustment of stress gradient along the FMF, the induced LPG can provide good $LP_{01}$-to-$LP_{11}$ mode conversion in the 1-μm spectral band with broad bandwidth of about 155 nm. The few-mode LPGs are then utilized in the generation of optical doughnut beams with a tunable ytterbium-doped fiber laser (YDFL).

A schematic diagram based on the stress-induced LPG is shown in Fig. 1. A few-mode fiber for the 1-μm spectral band (Corning SMF-28) is sandwiched between a periodical

V-grooved plate and a flat plate, where the pressure is applied for producing periodic stress on the FMF. The periodicity of grooves is 400 μm. There are three different light sources used in the experiment: an ytterbium-doped fiber laser (YDFL) for narrow-band doughnut generation, an amplified spontaneous emission (ASE) source for broadband doughnut generation, and an incandescent white-light. To generate and observe the doughnut beam, a home-made YDFL with wavelength tuning range of 1030.0 nm ~ 1097.8 nm is used as the light source. An ytterbium-doped fiber amplifier (YDFA) is used as the gain medium of YDFL. The output port of YDFA is connected to the input port through an output coupler and a tunable filter, thus forming a ring-cavity fiber laser. When the input of the YDFA is disconnected, the output of YDFA becomes an ASE source, which can be used to generate broadband doughnut beam.

When the pressures on both sides of the V-grooved plate, $P_1$ and $P_2$, are different, a stress gradient will be introduced along the fiber. Light from a wavelength-tunable ytterbium-doped fiber laser is launched into the single-mode fiber (SMF) and then coupled into the FMF with fusion splice. The SMF used for the 1-μm spectral band is Corning HI1060. The cutoff wavelength of the fundamental mode for SMF-28 is about 1260 nm, which means that SMF-28 is a two-mode fiber in the 1-μm band, supporting $LP_{01}$ and $LP_{11}$ modes. After tuning the polarization controllers, an optical doughnut beam can be generated and is recorded by a beam profiler. The transmission spectrum of the induced FMF-LPG is measured by using a white-light source, where the output port of FMF-LPG is connected to a mode stripper to remove the higher-order modes, and then characterized by an optical spectrum analyzer (OSA).

For a constant-period LPG generated by tuning the angle of the grooved plate against the axis of straight fiber, the effective periodicity $\Lambda_{eff}$ of the gratings can be expressed as:

$$\Lambda_{eff} = \frac{\Lambda_0}{\cos\theta} \qquad (1)$$

where $\Lambda_0$ is the periodicity of V-grooves and $\theta$ is the angle between the fiber and the normal of grooves. The resonance wavelength $\lambda_{res}$ and the effective grating periodicity $\Lambda_{eff}$ are related by the phase-matching condition:

$$\lambda_{res} = \Lambda_{eff}(n_{01} - n_{11}) \qquad (2)$$

where $n_{01}$ and $n_{11}$ are the effective indices of the $LP_{01}$ and $LP_{11}$ modes, respectively. Therefore, the resonance wavelength can be adjusted by the effective grating period or by the effective index difference. By adjusting the angle $\theta$ of the grooved plate, the grating periodicity $\Lambda_{eff}$ varies accordingly, and the resonance wavelength of LPG can be tuned.[12]

On the other hand, the effective index difference may be changed by photo-elastic or thermo-optic effects.[15,16] In addition, the induced birefringence caused by lateral stress will split the resonance wavelength into a shorter (blue-shift) and a longer (red-shift) resonance wavelengths. We should mention here that the effective grating periodicity of the stress-induced LPG is determined by the periodicity V-grooves, which is independent of the magnitude of lateral stress. As a contrast, when an LPG (fabricated by ultraviolet exposure, electric arc discharging, or $CO_2$ laser inscription) undergoes longitudinal force along the fiber axis or lateral stress between two flat plates, the grating period is subject to change, which will lead to the shift of resonance wavelength by an amount of

$$\Delta\lambda_{res} = \Lambda_0\Delta(n_{01} - n_{11}) + (n_{01} - n_{11})\Delta\Lambda_0 \qquad (2)$$

where $\Lambda_0$ is the original period of these LPGs, $\Delta(n_{01} - n_{11})$ and $\Delta\Lambda_0$ are the changes in effective index difference and grating period, respectively.

Figure 2 shows the measured transmission spectrum of LPG with $\theta = 30°$ and $\Lambda_{eff} = 462$ μm, for which the pressures $P_1$ and $P_2$ are different, and a stress gradient exists along the FMF. A broadband transmission dip can be observed around 1060 nm. We find that the transmission loss is approximately equal to or larger than 10 dB in the spectral range from 980 nm to 1135 nm, which confirms good $LP_{01}$-to-$LP_{11}$ mode conversion. As compared with the typical 10-nm bandwidth of conventional LPGs, our stress-induced LPG shows an ultra-broad 10-dB bandwidth of about 155 nm and 3-dB bandwidth of about 415 nm. The formation of this ultra-broadband LPG is different from that of Ref. 12, where the resonance bandwidth is broadened by the bending of optical fiber, leading to the change of effective period along the LPG. In our work, the LPG resonance located in the 1-μm band with much broader bandwidth, and the underlying mechanism might be attributed to the variation of propagation constant and effective refractive index difference along the FMF, which is a result of gradual fiber deformation produced by the stress gradient, as schematically shown in Fig. 3. We may claim that the stress gradient causes the dual effects of position-dependent photoelasticity and effective indices of core and cladding, modifying the effective index difference and resonance wavelength of the FMF-LPG and leading to an ultrabroad LPG bandwidth. Using Eq. (2), the effective index difference between the $LP_{01}$ and $LP_{11}$ modes is estimated to be from $2.12 \times 10^{-3}$ to $2.46 \times 10^{-3}$.

After proper adjustment of the polarization controllers, we have successfully generated a doughnut beam from the FMF. While tuning the YDFL wavelengths from 1030.0 nm to 1097.8 nm, we observe that the doughnut beams can be constantly generated without the necessity of LPG adjustment, although the cylindrical symmetry of doughnut can be optimized by slightly tuning the polarization controllers. Figures 4(a)-(c) show the recorded doughnut beams for the YDFL wavelengths of 1030.0 nm, 1060.0 nm, and 1097.8 nm. When the input of the YDFA is disconnected, we find that the ASE beam profile is still an optical doughnut, as shown in Fig. 4(d). Figures 5(a) and (b) show the typical output optical field distribution directly from the FMF without and with the LPG, respectively. The light source is the YDFL. By inserting a linear polarizer between the FMF and the beam profiler, the optical pattern becomes bowtie-shaped and rotates with the polarizer transmission axis correspondingly [Fig. 5(c)-(f)], which demonstrated that the doughnut beam is the superposition of orthogonally polarized $HG_{10}$ and $HG_{01}$ modes in the FMF. Therefore, by using this ultra-broadband LPG, optical doughnut beams with vector nature have been successfully generated, covering the entire spectral range of ytterbium-doped fiber lasers. By using a constant period V-grooved plate and a flat plate, the central wavelength and rejection bandwidth of stress-induced FMF-LPGs can be tuned over a broad spectral range with adjustable transmission loss. The method is simple and the in-situ adjustment is feasible for the generation of either narrow-band or broadband optical doughnut beams.

In this paper, we have made ultra-broadband LPGs in a few-mode fiber by using periodic mechanical stress provided by a constant-period V-grooved plate. The center wavelength of resonance band is tuned by the angle between the fiber and the grooves, while the bandwidth can be tailored by the stress gradient. With the adjustment of stress gradient along FMF, high efficient $LP_{01}$-to-$LP_{11}$ mode coupling occurs in the spectral range of 980 nm to 1135 nm. Wavelength-tunable optical doughnut beams have been successfully generated from 1030.0 nm to 1098.7 nm with a tunable YDFL source. After analyzing the polarization state of the optical beam with a linear polarizer, the vector nature of the generated doughnut beam has been confirmed. The stress induced mode converter has the potential to generate OAM beams by tuning the birefringence of fiber via the adjustment of average stress by the V-grooved plate, which might provide simultaneous control of polarization and phase of optical doughnut beams in the future.

**Acknowledgments**

We thank partially financial support by the Ministry of Science and Technology, Taiwan, under grants MOST 109-2221-E-845-002 and MOST 110-2221-E-845-004.

**Figure Captions**

**Fig. 1.** Schematic setup of few-mode LPG and the generation of optical doughnut beam.

**Fig. 2.** Transmission spectrum of the LPG.

**Fig. 3.** Schematic of cross-sectional deformation along the few-mode fiber.

**Fig. 4.** (a)-(c) Doughnut beams for the YDFL wavelengths of 1030.0 nm, 1060.0 nm, and 1097.8 nm; (d) ASE doughnut beam.

**Fig. 5.** (a)-(b) Beam profiles of the fiber output without and with few-mode LPG, respectively. (c)-(f) Patterns of the optical beams after passing the doughnut beam of (b) through a linear polarizer with different orientation of the polarizer transmission axis.

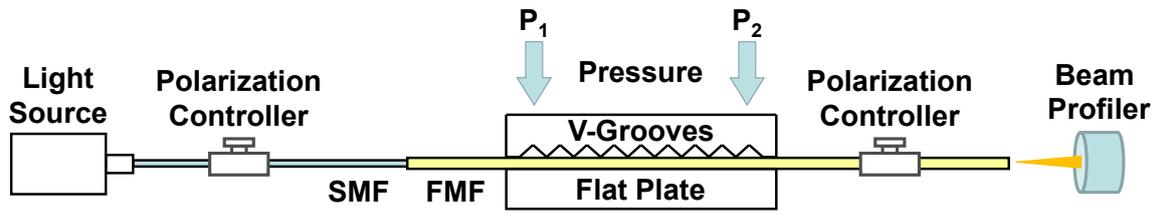

Fig. 1.

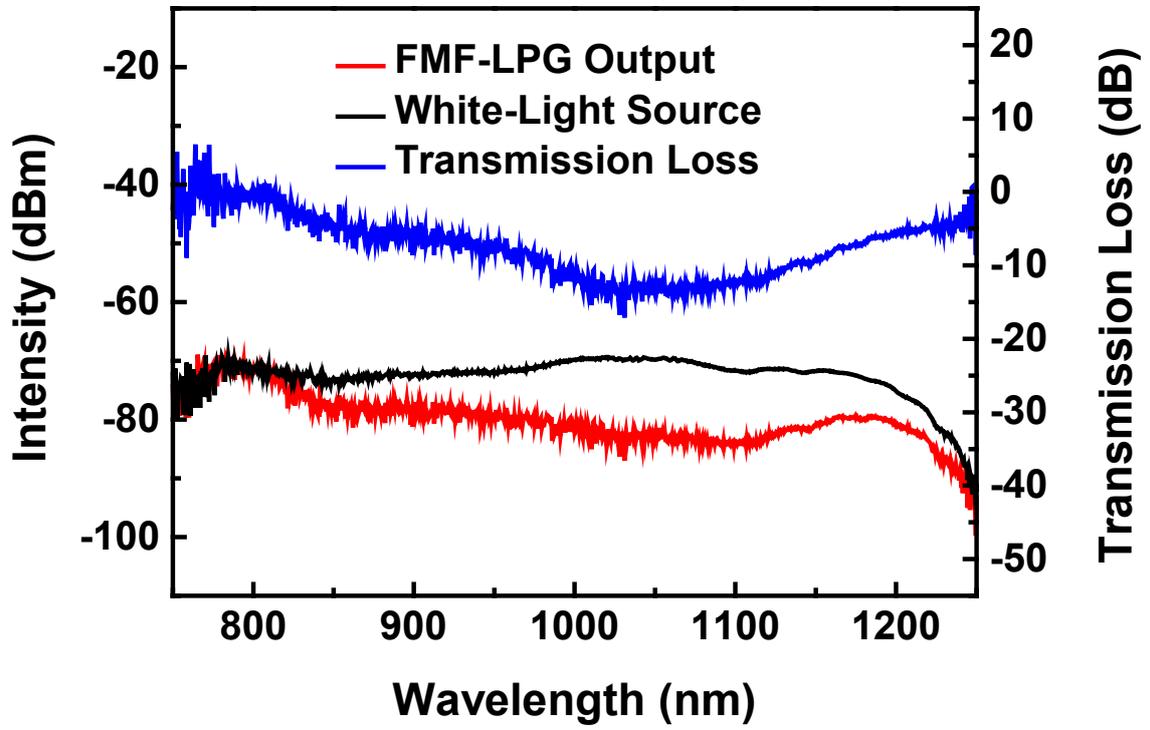

Fig. 2.

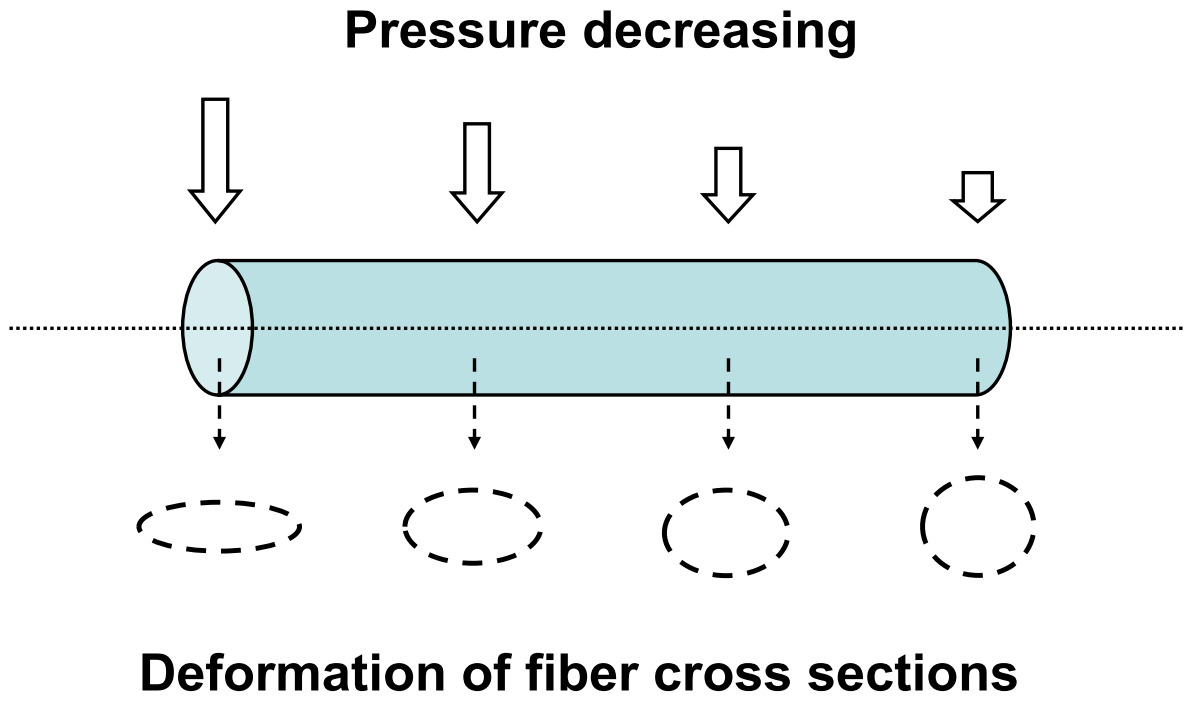

**Fig. 3.**

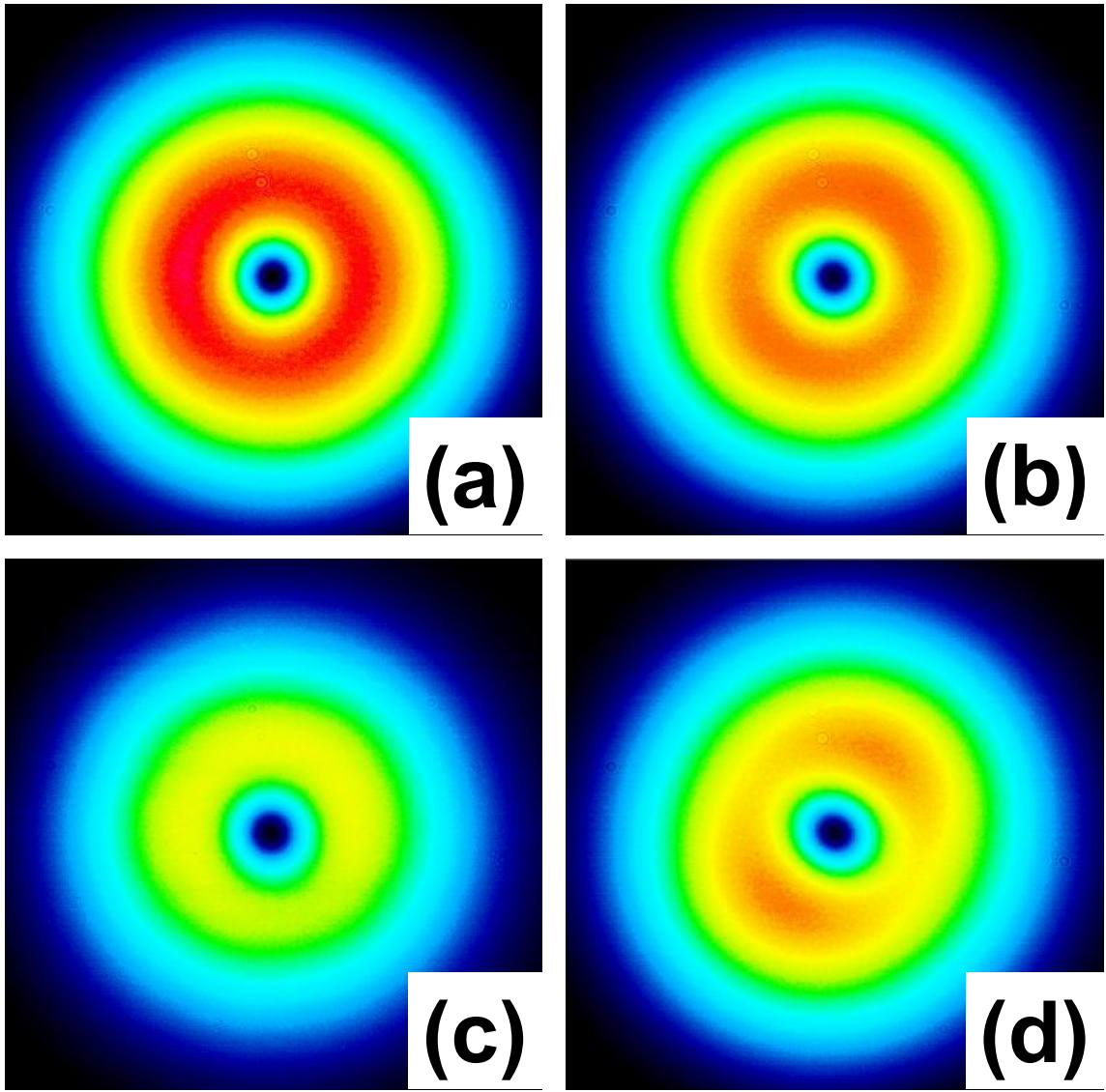

Fig. 4.

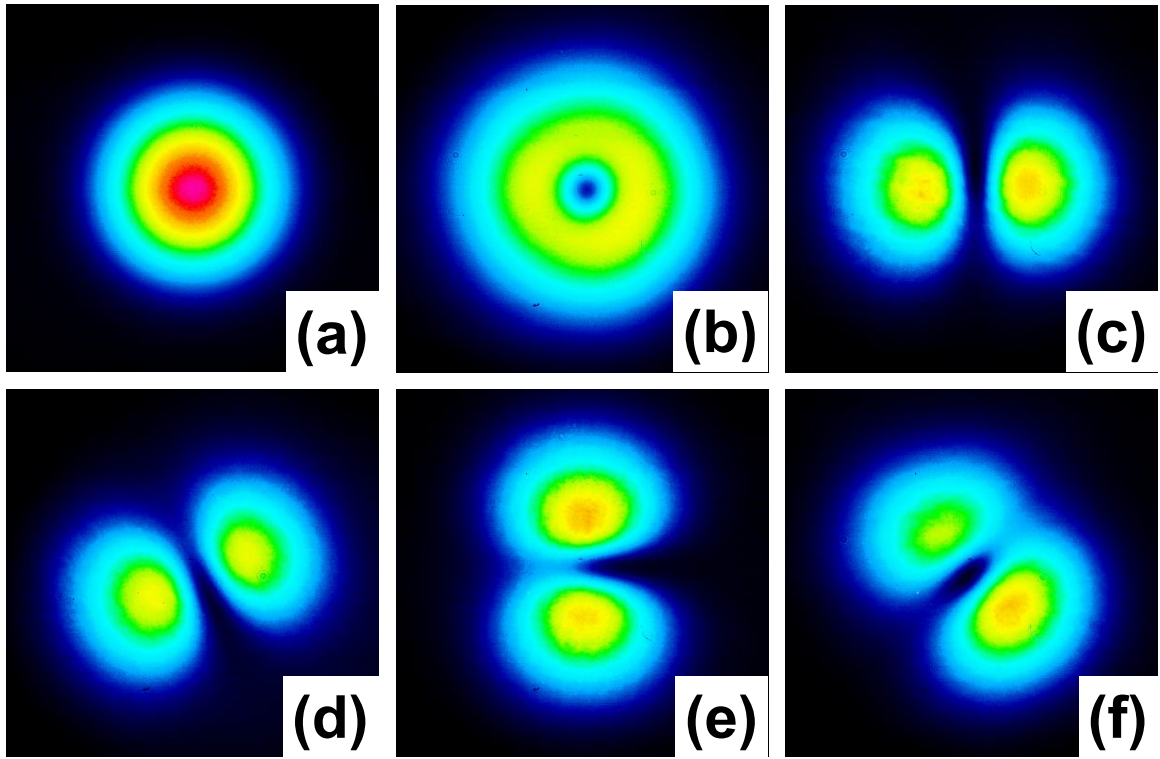

Fig. 5.